\documentclass[10pt,conference]{IEEEtran}
\IEEEoverridecommandlockouts
\usepackage[utf8]{inputenc}
\usepackage{cite}
\usepackage{amsmath,amssymb,amsfonts}
\usepackage{algorithmic}
\usepackage{color,soul}
\usepackage{graphicx}
\usepackage{textcomp}
\usepackage{xcolor}
\newcommand\mycaption[1]{{\footnotesize #1}}

\def\BibTeX{{\rm B\kern-.05em{\sc i\kern-.025em b}\kern-.08em
    T\kern-.1667em\lower.7ex\hbox{E}\kern-.125emX}}
\begin{document}

\title{Along the Margins: Marginalized Communities' Ethical Concerns about Social Platforms\\

}

\author{\IEEEauthorblockN{1\textsuperscript{st} Lauren Olson (she/they)}
\IEEEauthorblockA{\textit{Software and Sustainability, Social AI} \\
\textit{Vrije Universiteit Amsterdam}\\
Amsterdam, the Netherlands \\
l.a.olson@vu.nl}
\and
\IEEEauthorblockN{2\textsuperscript{nd} Emitzá Guzmán (she/her)}
\IEEEauthorblockA{\textit{Software and Sustainability} \\
\textit{Vrije Universiteit Amsterdam}\\
Amsterdam, the Netherlands \\
e.guzmanortega@vu.nl}
\and
\IEEEauthorblockN{3\textsuperscript{rd} Florian Kunneman (he/his)}
\IEEEauthorblockA{\textit{Social AI} \\
\textit{Vrije Universiteit Amsterdam}\\
Amsterdam, the Netherlands \\
f.kunneman@vu.nl}
}

\maketitle

\begin{abstract}

In this paper, we identified marginalized communities' ethical concerns about social platforms. We performed this identification because recent platform malfeasance indicates that software teams prioritize shareholder concerns over user concerns. Additionally, these platform shortcomings often have devastating effects on marginalized populations. We first scraped 586 marginalized communities' subreddits, aggregated a dataset of their social platform mentions and manually annotated mentions of ethical concerns in these data. We subsequently analyzed trends in the manually annotated data and tested the extent to which ethical concerns can be automatically classified by means of natural language processing (NLP). We found that marginalized communities' ethical concerns predominantly revolve around discrimination and misrepresentation, and reveal deficiencies in current software development practices. As such, researchers and developers could use our work to further investigate these concerns and rectify current software flaws.

\textit{General Abstract---}
In this paper, we identified marginalized communities' ethical concerns about social platforms. We did this because recent platform wrongdoing indicates that software teams prioritize profit over user concerns. Additionally, these platform shortcomings often have devastating effects on marginalized populations. To accomplish this, we collected Reddit posts from marginalized communities' subreddits where users mention social media platforms. Then, we labeled whether posts contained mentions of ethical concerns, like privacy or misinformation. Finally, we established trends within the resulting data and used artificial intelligence (AI) to find these ethical concerns automatically. We discovered that marginalized communities' ethical concerns revolve around discrimination and misrepresentation, among other problems, and reveal deficiencies in current social platforms. As such, researchers and software engineers could use our work to further investigate these concerns and rectify present software flaws.
\end{abstract}

%abstracts should answer:
  %1. What did we do
    
  %2. Why did we do it

  %3. How did we do it

  %4. What did we find

  %5. What do we think this means


\section{Introduction}
\label{sec:intro}
As society transitions online, inequities in the physical world are encoded into the digital world. Ethical concerns, such as censorship, misinformation, and discrimination, are common in today's software products. For example, just in the past few years, software platforms such as Facebook and Instagram have played a role in the censorship of Palestine's protests of Israel's forced eviction~\cite{biddle_2022} and the spread of misinformation and hate speech during Myanmar's genocide of the Rohingya people~\cite{rohingya_2022}. %For example, in May 2021, during Palestinian protests of their forced eviction, Meta's content moderation practices on Facebook and Instagram led to censorship of protesters' documentation of Israeli forces' violence~\cite{biddle_2022}. After Israel's conduct during this conflict, the International Criminal Court brought war crimes charges against Israel~\cite{federman_2021}. Another example of Meta profiteering from the conflict of other countries is Facebook's involvement in Myanmar's genocide of the Rohingya people. By allowing the spread of misinformation and hate speech, and failing to restrict government officials' propaganda, they played a role in the rape and murder of thousands of people~\cite{rohingya_2022}. 
This problem of misinformation and hate speech causing real-world violence is also relevant in other contexts. Acts of violence in the US, including mass shootings, indicate that echo chambers and hate speech online contribute to the systematic dehumanization of women, queer people, migrants, and racial minorities, with perpetrators radicalized on Discord\cite{chayka_2022}, Reddit\cite{woolf_2014}, Twitter\cite{digital_hate2022}, Instagram\cite{tiffany_2020}, and Facebook\cite{digital_hate2022}, among other platforms. %To combat platforms' malfeasance, we will focus on making users' ethical perspectives on software more usable for developers. 

Billions of these marginalized users are at the mercy of development decisions made by a small demographic of the world population. Most software developers who curate these platforms are white, middle to upper class, cisgender, heterosexual, English-speaking men from the United States~\cite{costanza2020design}. Studies show that developers' political affiliations affect their design decisions~\cite{costanza2020design}. However, this problem cannot be entirely solved by hiring software developers from diverse backgrounds. Even when teams are diverse, development decisions tend to reflect shareholder concerns, rather than user concerns~\cite{costanza2020design}. 

To recenter the focus of today's software development from privileged to marginalized communities, we aggregate marginalized users' feedback on software. In doing so, we can capture their perspectives on current ethical issues. Although we cannot restructure economic systems of power, we can at least make this information more accessible to developers. While previous attempts to collect user feedback have a population bias of mostly middle-aged men~\cite{tizard2020voice}, we focus on unheard voices. We gather users' ethical concerns related to software platforms from Reddit to determine the concerns and desires of systematically marginalized users, which are not represented during software development. We chose Reddit as our data source because of its high character-limit, audience-specification, and anonymity features. Specifically, we aggregated a large dataset of their perspectives on software, manually analyzed a sample of their ethical concerns, and developed models to process and understand their concerns on a large scale to make these viewpoints available to developers and researchers.

This work contributes to (1) the field of user feedback for software evolution by disaggregating feedback from privileged idenities and (2) design justice work by considering a data-driven process to supplement user studies. Although previous research has considered differences in user feedback concerning gender\cite{gender_user}, language\cite{same_same}, and culture\cite{user_cross_cultural}, this paper is the first to look at differences in feedback concerning ability, sexuality, gender identity, sex, race, and socio-economic status (SES). Furthermore, our work is the first to include intersectional groups, groups with intersecting marginalized identities, whose concerns cannot be accurately represented by considering identity on a one-dimensional scale. In addition, design justice work typically features small-scale user-centered studies, which are critical for accurately representing users' concerns. However, by collating marginalized group feedback on platforms and identifying common trends, our research could allow other researchers greater insight into their users' concerns at the start of their studies' designs to allow for deeper and more targeted lines of questioning. 

\section{Research Design}
The main goal of this study is to collect and analyze rich user feedback from marginalized communities detailing their ethical concerns about software. Furthermore, we aim to obtain feedback containing not just ethical concerns but context and detail on their experiences to allow developers and researchers to understand marginalized communities' feedback and act on it.

\subsection{Data Source}
\label{sec:reddit}
We use Reddit as our data source because of its long-form posts, subreddit structure, and throwaway accounts. A main challenge in eliciting user feedback is the lack of actionable content \cite{context_martens}. Previous studies have focused on social media sites like Twitter, which restricts character length to 280 characters, as well as the Apple App Store and Android Google Play Store, which cap reviews at 6,000 and 4,000 characters, respectively. Reddit, in comparison, limits posts to 40,000 characters. This encouragement of lengthy discussion by design makes Reddit a potentially rich source for user feedback, making posts more likely to include context surrounding their experiences with the platforms. 

Beyond Reddit’s lengthy character count, another unique feature, compared to other social media platforms, is the structuring of its content. Instead of subscribing to content from specific accounts, which typically represent one user or entity, users subscribe to \textbf{subreddits}, which represent groups of users. The topic of the subreddit unites these groups; for example, some subreddits include “r/funny,” “r/movies,” and “r/amsterdam.” This structure allows users to filter their feeds to only topics and populations of interest. In addition, these subreddits are communities with content moderators and rules of behavior. This user-led moderation gives users greater control over the users they interact with and the content they view. This structuring thus allows the explicit allocation of platform space for groups who may get overlooked through other social media platforms’ use of popularity measures to filter and promote content. For marginalized communities who often face hate in less regulated online spaces, these subreddits can provide a safe(r) space to find others with similar situations ~\cite{workmanfront}. For example, women in r/rapecounseling can find an otherwise non-existent outlet for being heard and believed ~\cite{o2018today}. Because of this audience-specifying feature, we expect marginalized communities to be more willing to share ethical concerns relating to their marginalized identities. 
 
Additionally, Reddit allows users to anonymize their posts through the use of \textbf{throwaway accounts}, further ensuring that those with sensitive information can feel safe posting. For example, Leavitt et al. found “that women are much more likely to adopt temporary identities than men”~\cite{leavitt2015throwaway}. Leavitt et al. posit that this adoption of throwaway accounts is motivated by users’ fear that people in their real lives might uncover their identities. 
 
\subsection{Research Questions}
This study focuses on three main research questions: \\
\indent \textbf{(RQ1)} Is Reddit a fertile source of feedback on ethical concerns about software from marginalized communities?

\textbf{(RQ2)} What types of ethical concerns regarding software do marginalized communities have on Reddit?

\textbf{(RQ3)} What is the automation potential of extracting and classifying ethical concerns in Reddit posts from marginalized communities?

\section{Related Work}
\subsection{Marginalized Communities}
Previous studies designed to capture marginalized communities' software preferences have conducted user-centered studies with small groups of marginalized communities. For example, a less recent study by Koepler et al. intended to collect the perspectives of homeless people on social media by surveying 199 people on their preferences \cite{koepfler2013stake}. They surveyed these users to determine which platform features could aid homeless users in sharing resources and forming communities online. 

More recently, user-centered design has sought to give the user a more active role in the development process. For example, De Vito et al. performed an interactive study to discover 31 queer users' design values to combat stigmatization and harm online. Also, Rankin et al. discovered 10 black youths' different persona and utility conceptualizations of a Siri-like agent during 7 interactive design sessions ~\cite{rankin2021resisting}. These two more recent studies are samples of critical work in disaggregating development to consider how marginalized users' concerns may differ from shareholders'. 
These studies are crucial and could be supplemented by our collected data, content analysis, and NLP models as our data could aid in forming user-centered studies' design activities and questionnaires to deepen and facilitate these studies. Our study, in contrast, aggregates the perspectives of thousands of users and 35 marginalized communities. In addition, ours is the first to analyze marginalized groups' ethical concerns from social media.

\subsection{User Feedback}
\label{sec:feedback}
Previous work found that user feedback is essential for software quality and identifying areas of improvement~\cite{Pagano2013}. 
\twocolumn{
\begin{table*}
\label{table:subreddits}
 \caption{Subreddits Chosen for Manual Annotation}
\begin{tabular}{|l|l|l|l|l|l|l|}
\hline
\textbf{Physical Disability} & \textbf{Neurodivergent} & \textbf{LGBTQIA+} & \textbf{Race} & \textbf{Global South} & \textbf{Women\slash AFAB} & \textbf{Lower SES}   \\\hline
prostatitis                  & narcissism                                         & actuallylesbian                        & asianmasculinity                      & MalaysiaPolitics                              & whereAreTheFeminists                   & almosthomeless   \\\hline

disabledgamers               & adhd                                               & asexualdating                          & southasianmasculinity                 & asklatinamerica                               & femalefashionadvice                    & homeless         \\\hline

sciatica                     & mentalhealthsupport                                & meetlgbt                               & indianSkincareAddicts                 & askthecaribbean                               & askfeminists                           & homelesssurvival \\\hline

disabled                     & malementalhealth                                   & transadoption                          & aznidentity                           & beautytalkph                                  & girlsgonewired                         & vagabond         \\\hline

spinalcordinjuries           & \textit{mentalhealthPH}                                & honesttransgender                      & \textit{BlackGirlDiaries}                     & MalaysianPF                                   & \textit{TwoXIndia}                              & vandwellers   \\\hline  
\end{tabular}
\end{table*}
}
With the rise of mobile applications and social media, research proposed to elicit feedback from crowds of geographically distributed users~\cite{groen2017crowd} and called for the mass participation of software users during different stages of software development\cite{Johann2015}.
Pagano and Maalej~\cite{Dennis2013}, and Hoon~\cite{hoon2013analysis} were among the first to study user feedback in app stores.
They performed exploratory studies and found that this platform contains valuable information for software evolution. In a similar line, other work \cite{Guzman2016, guzman2017little, nayebi2018app, williams2017mining} found that user feedback on Twitter also contains valuable information for software evolution. 

Recent work~\cite{iqbal2021mining} studied posts about specific software applications on Reddit and found that this platform is also a good source of user feedback when evolving software. Previous work also found that there are cultural differences in how feedback is given~\cite{user_cross_cultural,fischer2021does} that more men give feedback about software~\cite{tizard2020voice,gender_user} and that the majority of these are in the 35-44 age range~\cite{tizard2020voice}. 

There are few studies which focus on ethical concerns in user feedback. Tushev et al.~\cite{tushev2020digital}, Besmer et al.\cite{besmer2020investigating}, Li et al.\cite{reddit_privacy}, and Khalid et al.\cite{khalid2014mobile}, all consider either discrimination~\cite{tushev2020digital} or privacy\cite{besmer2020investigating},\cite{reddit_privacy},\cite{khalid2014mobile} in user feedback on software. In addition, Shams et al.~\cite{shams2020society} and Obie et al.~\cite{obie2021first} analysed human values violations in app reviews with the Schwartz theory of basic values~\cite{schwartz2012overview}, which includes 11 values. Although, this theory and its values were originally developed for the psychology field so its values don't align closely with software. However, later work created an ethical concerns taxonomy specifically developed for user feedback on software platforms \cite{ethical_concerns}. We use this taxonomy for part of manual analysis (see Section~\ref{sec:finergrained_labeling}).

To our knowledge, no research has studied how marginalized communities give feedback about software and which ethical concerns these groups have about the software they use. We address this gap in our work.

%this section is intended to be about the gaps is user feedback and why that necessitates this paper 
%- how twitter and the app store may be limited due to structure (character limit, audience) and our inability to disaggregate certain data - we can only look at some demographic information (country of origin?), genderize.io isn't completely accurate

%- Mining reddit as a new source for software requirements

%* in tahira's paper, she finds that reddit is a good source of feedback, but she looks in the platform related subreddits, which likely has the same vulnerabilities 

%- Voice of the users: A demographic study of software feedback behaviour \cite{tizard2020voice}
%also, we *know* app store is men mostly 35-44, so it really doesn't represent everyone

% - How Do Users Like This Feature? A Fine Grained Sentiment Analysis
% - How can i improve my app? Classifying user reviews for software maintenance and evolution
% - A Needle in a Haystack: What Do Twitter Users Say about Software?
% - User Feedback in the App Store: A Cross-Cultural Study
% - Gender and User Feedback: An Exploratory Study \cite{gender_user}
% - Same Same but Different: Finding Similar User Feedback Across Multiple Platforms and Languages

% - A first look at human values-violation in app reviews

% - Digital discrimination in sharing economy a requirements engineering perspective \cite{tushev2020digital} -> only looked at tweets, hard to understand context, higher character length could make posts more actionable, 

\section{Research Method}
Our research method aims to collect marginalized communities' Reddit posts, identify and classify ethical concerns regarding software platforms within these posts, and set up our posts for automated identification and classification of ethical concerns. First, we \textit{select seven marginalized communities} to focus on, search Reddit for their subreddits, and \textit{scrape these subreddits,} resulting in a dataset of 459,523 posts. Then, we \textit{select the software platforms} to focus on and search for these platforms' names within the scraped Reddit posts. This filtering produces a dataset of 23,533 posts that mention platforms. Next, we \textit{manually annotate a sample} of 2,201 posts to identify ethical concerns within these posts. After identifying all ethical concerns within this sample, we classify each post's ethical concern by type. Next, to \textit{prepare for the process of automatically identifying and classifying ethical concerns,} we trained a binary and multi-class classifier with our previously annotated data. In the following, we describe each of these steps in greater detail.

\subsection{Marginalized Community Selection} Unless developers are told otherwise, they assume users are ``white, male, abled, English-speaking, middle-class US citizens"~\cite{costanza2020design}. Even if these biases are unintentional, or corrected by diverse team members, because purchasing power is held by these same privileged groups, platforms structure their development around these groups' priorities. To counter this bias, the seven demographic groups we consider are (1) race (2) women/AFAB, (3) LGBTQIA+, (4) physically disabled, (5) neurodivergent,  (6) lower socio-economic status (SES), and (7) Global South.

To create a comprehensive list of subreddits that cover a broad range of subgroups within the chosen marginalized communities, we used Reddit's directory\footnote{https://www.reddit.com/subreddits/a-1}. First, we scanned the list of subreddits alphabetically, searching for those related to our seven groups. Then, we checked the description for the subreddit to ensure that the subreddit was accessible and related to the perceived category. For example, the “r/chad” and “r/haiti” subreddits were private, and the “r/fiji” subreddit is a subreddit dedicated to a fraternity, not the country. Finally, we included subgroups for the more extensive categories as a general criterion. For example, we included both the subreddit for Tunisia, “r/tunisia,” as well as “r/tunisianjobs,” a subreddit for Tunisians looking for jobs; however, we did not include “meme” or other image-related subreddits as the majority of posts did not contain text-based content. 

Next, we describe more in detail how we found the subreddits for each of the marginalized communities considered in this study.

\subsubsection{Race}
For the race category, any mentions of race within the title of the subreddit warranted inclusion in this category. For ease of identification and due to Reddit's majority US user base, we used the US' official racial categories, ``White, Black or African American, American Indian or Alaska Native, Asian, and Native Hawaiian or Other Pacific Islander," as well as any sub-categories. 

\subsubsection{Women/AFAB}
This category intends to cover the perspectives of those who receive discriminatory treatment for either (1) being perceived as a woman or (2) owning a female body. As such, we included feminism, women, and assigned female at birth (AFAB) related groups. For example, we included groups with “TwoX” in the title, which is a reference to an AFAB’s genetic composition of two X chromosomes, subreddits related to pregnancy and breastfeeding, or “feminism,” “wom[e/a]n,” “female” in the title. It was necessary to check these subreddits as many groups that may seem for women are groups that post and rate photos of women (e.g., “r/latinas”). 

\subsubsection{LGBTQIA+}
In this paper, we use queer as an umbrella term for the LGBTQIA+ community. To generate a list of LGBTQIA+-related subreddits, we searched for subreddits relating to each letter of the acronym, adding relevant recommended communities as well, like ``r/agender," ``r/abrosexual," and ``r/Crossdressing\_support." 

\subsubsection{Physical disability}

 The World Health Organization defines \textit{disability} as ``the interaction between individuals with a health condition...and personal and environmental factors;" \cite{who} as we cannot account for personal and environmental factors, we include all health-related subreddits.
 %This criterion includes subreddits for symptoms of debilitating medical conditions; for instance, we included the subreddit for dermatillomania, also known as skin picking, which is a symptom of many anxiety-related conditions. 

\subsubsection{Neurodivergence}
The requirements applied to the physical disability category are the same as this category. All subreddits relating to neurodivergent conditions are included. 

\subsubsection{Lower SES}
For the lower SES category, we included subreddits related to homelessness or near homelessness.

\subsubsection{Global South}
To develop the group of Global South-related subreddits, we referenced a list of Global South countries\footnote{https:\/\/worldpopulationreview.com\/country-rankings\/global-south-countries} to determine whether a subreddit was a mention of a Global South country. 

This marginalized community selection process resulted in a comprehensive list of 586 subreddits. The community with the largest number of subreddits is the Global South, with 216 subreddits, and the smallest community by far is lower SES, with only seven found subreddits.

\subsection{Data Scraping}
We scraped the posts of our gathered subreddits with an existing scraper\footnote{https://github.com/JosephLai241/URS}. We scraped the Reddit posts on November 19, 2021. For each subreddit, we scraped until we collected either the entire subreddit's posts or the most recent 1,000 posts. The average collection period per subreddit is 5.8 years, making the average earliest post from December 2015. The complete list of subreddits is available in the replication package\footnote{https://doi.org/10.5281/zenodo.7194259}. This data scraping resulted in a dataset with 459,523 posts. Again, the Global South was the largest category, with 158,590 collected posts, and the lower SES category was the smallest, with only 4,517.

%\begin{figure}
 %   \caption{Subreddits by Category}

  %  \centerline{{\includegraphics[width=.5\textwidth]{figures/subreddits.png}}}%
    
   % \label{fig:subreddits}
%\end{figure}

\subsection{Platform Selection}
In this study, we focus on posts mentioning specific social software platforms. We chose to focus on social software platforms because these platforms tend to have frequent claims of platform malfeasance, as referenced in the Introduction ~\cite{biddle_2022},\cite{rohingya_2022},~\cite{chayka_2022}, \cite{woolf_2014},\cite{digital_hate2022},\cite{tiffany_2020}. We selected social platforms with at least 100 million users\footnote{https\:\slash \slash en.wikipedia.org\slash wiki\slash  List\_of\_social\_platforms\_with\_at\_least \\
\_100\_million\_active\_users\#cite\_note\-\ 1\-1}. We decided on this threshold to make it more likely to find enough posts per platform since our collected subreddits are not specifically dedicated to software platforms.  

\begin{figure}
    \caption{Number of Posts by Platform}

    \centerline{{\includegraphics[width=.5\textwidth]{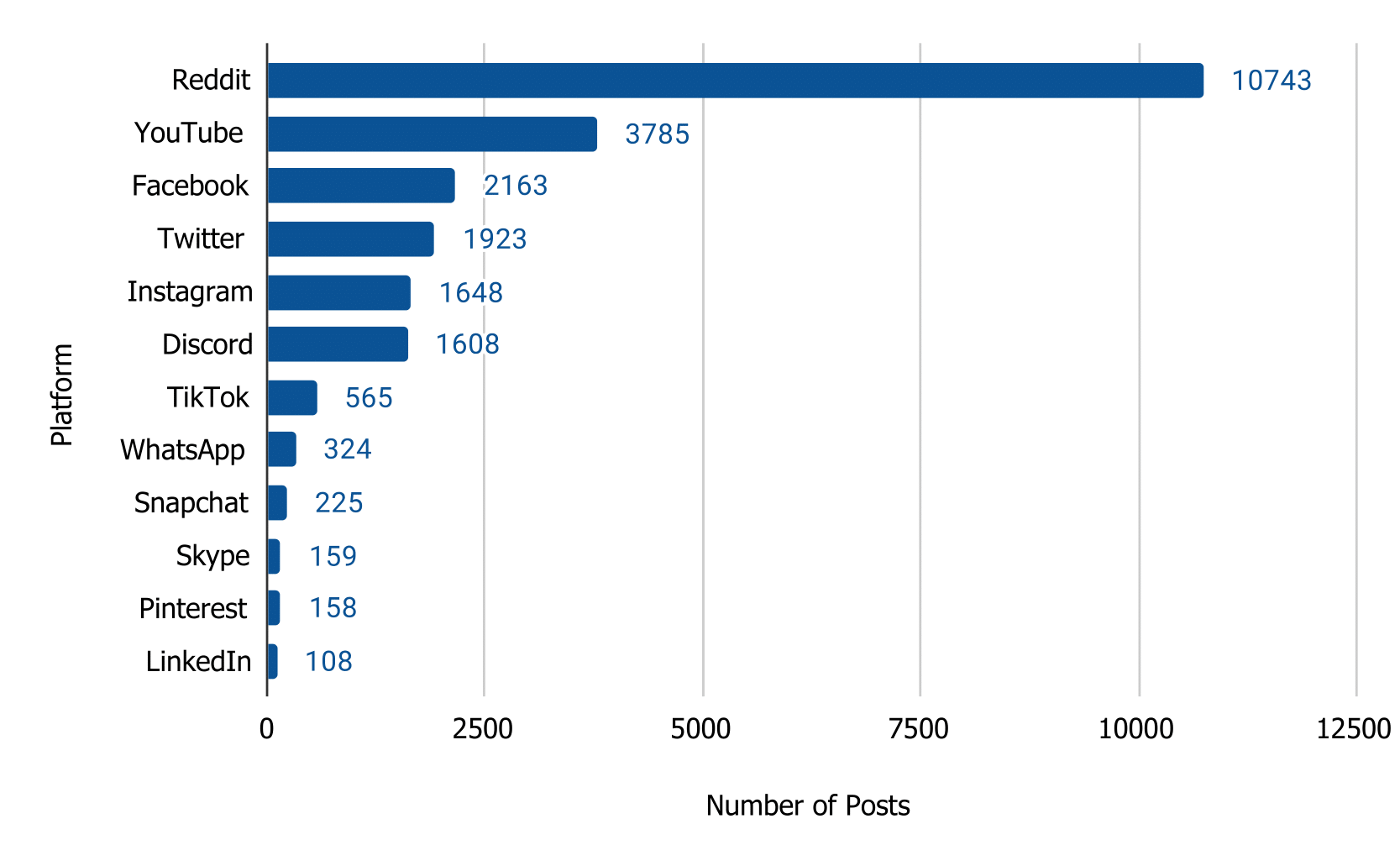}}}%
    
    \label{fig:mentions}
\end{figure}

After searching our dataset for mentions of these platforms, we found 23,533 posts. We removed all platforms with less than 100 posts in the dataset. We also removed platforms with names ``imo" and ``Line," as their mentions were likely to be references to the anagram “in my opinion” or the word “line.” In total, we included 12 popular social platforms in our study. We show them and their number of posts in Figure~\ref{fig:mentions}. Reddit had the greatest number of posts due to self-references at 10,743.
%These include WeChat, Douyin, Telegram, QQ, Weibo, QZone, Kuaishou, Tieba, Viber, Teams, PicsArt, and Likee. 
%\begin{figure}
 %   \caption{Total Posts}

  %  \centerline{{\includegraphics[width=.5\textwidth]{figures/posts.png}}}%
    
   % \label{fig:posts}
%\end{figure}

\subsection{Sampling}
Because we could not manually label over 20,000 posts, we needed to create a sample representing the dataset as accurately as possible. To create this sample, we scored each subreddit based on its number of platform mentions, with the intuition of capturing the most salient phenomena. We included the five highest-scoring subreddits in each category within the sample used during the annotation since we wanted to pay equal attention to each of the marginalized communities. We show the used subreddits in Table~\ref{table:subreddits} (intersectional groups are italicized). Next, we calculated the sample size for a 95\% confidence level, collecting the posts randomly from an aggregated pool of the chosen subreddits for each category. The resultant sampling accumulated 2,201 posts for annotation.

\subsection{Manual Annotation Setup}
\subsubsection{Ethical Concern Identification}
The annotation task was to identify whether ethical concerns were present in posts mentioning software platforms. To ensure that the task and definitions were clear, we initially performed six trial rounds of coding. Every round had two posts from each of our marginalized communities, adding up to 14 posts per round. Each round had only 14 posts due to our dataset's average post length of 962.27 characters, or around 150-200 words for each post. All rounds were completed with the first author and one other author, and the final four were performed with all three annotators. For the first round of annotation, annotators achieved a Krippendorf's alpha of .381. This initial score reflects this task's difficulty, mainly because most Reddit posts' intent was not to give direct feedback on ethical concerns about software. To correctly label ethical concerns, the annotators used an extensive annotation guide\footnote{https://doi.org/10.5281/zenodo.7194259} with definitions and examples\footnote{From actual Reddit posts.} detailing how to interpret and classify the posts. Our utmost concern in the process was avoiding the dismissal of more implicit ethical concerns.

\begin{figure}
   \caption{Manual Annotation Complexity}

    \centerline{{\includegraphics[width=.5\textwidth]{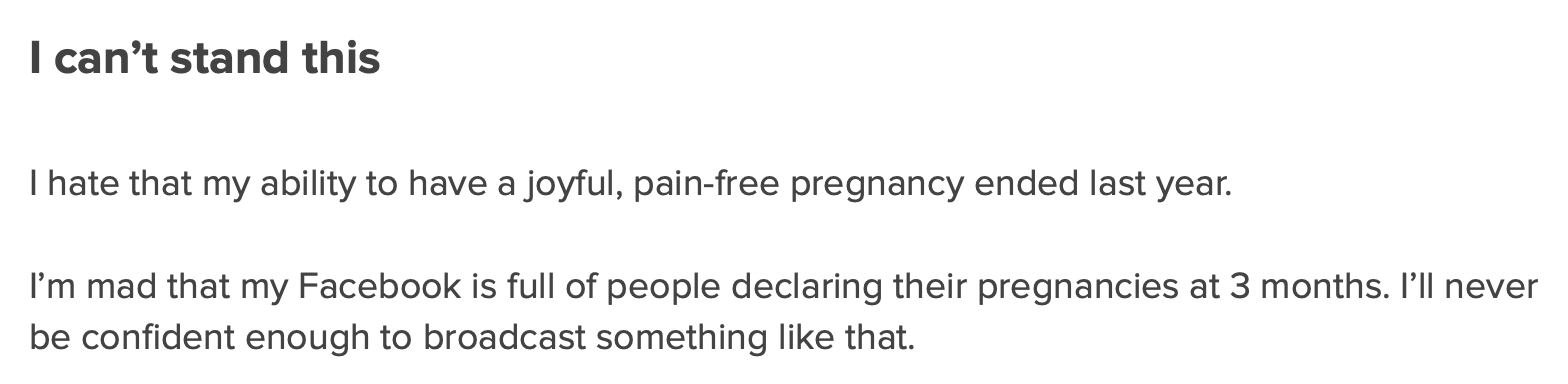}}}
    \mycaption{\textit{Post modified to protect author's privacy.}}
    \label{fig:example}
\end{figure}

As the rounds proceeded, we updated this annotation guide to clarify disputes between annotators. To simplify the labeling process, annotators decided to assume the users' correctness, platforms' responsibility and interpret ethical concerns as any ``worry or care the user or their group faces about what is right or good, that is not simply a bug report or feature request" on the platform. For example, in the post\footnote{https:\slash \slash www.reddit.com\slash r\slash PregnancyAfterLoss\slash comments\slash iux55d\slash \\ i\_hate\_this\slash} shown in Figure \ref{fig:example}, two points of ambiguity were raised by annotators. First, in our posts, often users raise an issue faced on a platform, but do not explicitly blame the platform for the issue. So, we decided that, regardless of user perception of the platform's role in the issue, we would classify these implicit ethical concerns as ethical concerns. Second, it is unclear whether the platform would be able to solve the user's concern. Again, we made the assumption that platforms were ultimately responsible for their users' experiences. As such, we categorized the example shown in Figure \ref{fig:example} as an ethical concern. After the three annotators achieved a Krippendorf's alpha of $1.0$ in the final round, the first author annotated the rest of the 2,201 posts on their own. After completing the initial identification of ethical concerns, we had a labeled dataset with 580 ethical concerns. 

\subsubsection{Ethical Concern Classification}\label{sec:finergrained_labeling}
To enable a more detailed analysis of these ethical concerns, the first author did a second round of labeling to categorize each of the ethical concerns into more specific labels like ``privacy," ``misinformation," etc. Before this step, the first and second author performed two trial runs with the labels used in this detailed analysis. In their final run, they scored a Cohen's kappa of .77. For this step, we used an existing ethical concerns taxonomy for software applications~\cite{ethical_concerns}. Table \ref{table:taxonomy} shows the categories of this taxonomy; the term definitions are available in our replication package\footnote{https://doi.org/10.5281/zenodo.7194259}. This taxonomy was supplemented with an ``other" category for ethical concerns that did not fit into any of the existing categories. This category introduces the capability to capture ethical concerns specific to marginalized communities and newly emerging ethical concerns. We also added a ``social isolation" category which was removed from the original taxonomy~\cite{ethical_concerns} due to lack of relevance, but was present within the current data, likely due to the different outlets for posting and our focus on marginalized communities. %As mentioned in Section \ref{sec:reddit}, the Google Play Store's shorter character limit and audience of customer support professionals makes it a rich source for feedback relating to feature requests and bug reports, but less so for ethical concerns relating to technology. 

\begin{table}[hbt!]
\centering
\caption{Ethical Concerns' Taxonomy}
\begin{tabular}{|c|c|c|}
\hline
\multicolumn{3}{|c|}{\textbf{Ethical Concerns}}  \\\hline

Accessibility & Harmful Advertising & Discrimination\\\hline

Addictive Design & Identity Theft & Scam\\\hline

Censorship & Inappropriate Content & Social Isolation*\\\hline

Content Theft & Privacy & Misinformation\\\hline

Cyberbullying & Safety & Other*\\\hline

\end{tabular}
\\
\mycaption{\textit{*Added categories}}
\label{table:taxonomy}
\end{table}

\subsection{Automated Extraction and Classification Setup}
The steps described in this section prepare our data to act as input for NLP classifiers, so they can detect and classify ethical concerns in Reddit posts, allowing software developers to more easily find ethical concerns and categorize them within Reddit posts.  
We cleaned the data as an initial step to create performative NLP models. First, we preprocessed the data to remove noise 
and allow for the comparison of similar words. Then, as part of the training phase, we balanced the data to ensure that our models were not biased towards one label due to their frequency, allowing them to classify based on salient features of the textual data. 

\subsubsection{Preprocessing} 
We first preprocessed the dataset using the NLTK\footnote{https://www.nltk.org/} library. We eliminated special characters and symbols, performed lowercasing and tokenization, eliminated stopwords using NLTK's English stopwords set\footnote{https://gist.github.com/sebleier/554280}, and stemmed the resulting words. 

Next, we vectorized the words using the \textit{TfidfVectorizer} with an n-gram range of 2. Finally, we also decreased max\_df to .5, to reduce reliance on Reddit-specific terminology, and sublinear\_tf to True, which weakens term frequency. 

In addition, due to the length of Reddit posts, we implemented a window algorithm, which only included the sentence mentioning the platform itself, along with one sentence before and one sentence after this sentence, thereby removing noise from the posts. 

\subsubsection{Handling Data Imbalance}
\label{sec:smote}
For the binary classification task, there were less ethical concern platform mentions than non-ethical concern platform mentions, signalling a slight data imbalance. In total, 26.49\% of the labelled reviews mentioned an ethical concern. Again, for the multi-class classification task, our distribution ethical concerns were unbalanced as well, with discrimination (25.88\%), censorship (24.39\%), other (10.06\%), cyberbullying (9.50\%), and addictive design (7.64\%) all at varying percentages.  To fix this, we implemented stratified 10-fold cross-validation and Synthetic Minority Oversampling Technique (SMOTE). Cross validation is a resampling method which trains and tests the models on discrete sections of the data, to decrease the models' dependence on any one section of the dataset. SMOTE oversamples minority classes by fabricating data points~\cite{chawla2002smote}. We only applied SMOTE on our training data. 
 

\section{Results}
This section addresses the frequency of ethical concerns within our marginalized communities' subreddits. We disclose general measures and the frequencies of ethical concerns within our marginalized communities and platforms. Next, we perform a content analysis on each type of ethical concern. Lastly, we report the automation potential of identification and classification of ethical concerns. 

\subsection{Frequency of Ethical Concerns Feedback}
\label{sec:manual}
The selected software platforms were mentioned in 5.1\% of the posts, 26.49\% of these posts expressed ethical concerns about software. These posts were relatively long, with an average length of 911.8 characters, or 140-182 words. However, although a post mentions a platform, the main subject of the post might not be the platform. This leads to swaths of irrelevant data within the posts. 

%\subsubsection{Popularity}
% On Reddit, \textit{score} is akin to the number of likes a post receives. Ethical concerns' popularity measures can be viewed in Figure~\ref{fig:pop}. In our dataset, ethical concerns have a mean score of 277.89, while those without have a mean score of 229.83. Indeed, the median score of ethical concern posts is double that of those without, a score of 46.00 as compared to 23.00.  
%Another measure of popularity to be considered is the number of comments that post generated. Again, our ethical concern posts thus have greater popularity than those without, with an mean average of 56.28 and 44.99 comments, respectively. 

%\begin{figure}
   % \caption{Ethical Concerns' Popularity}

    %\centerline{{\includegraphics[width=.5\textwidth]{figures/popularity.pdf}}}
    
   % \label{fig:pop}
%\end{figure}

\subsubsection{Marginalized Communities}
\begin{figure}
    \caption{Ethical Concern Frequency by Marginalized Group}

\centerline{{\includegraphics[width=.5\textwidth]{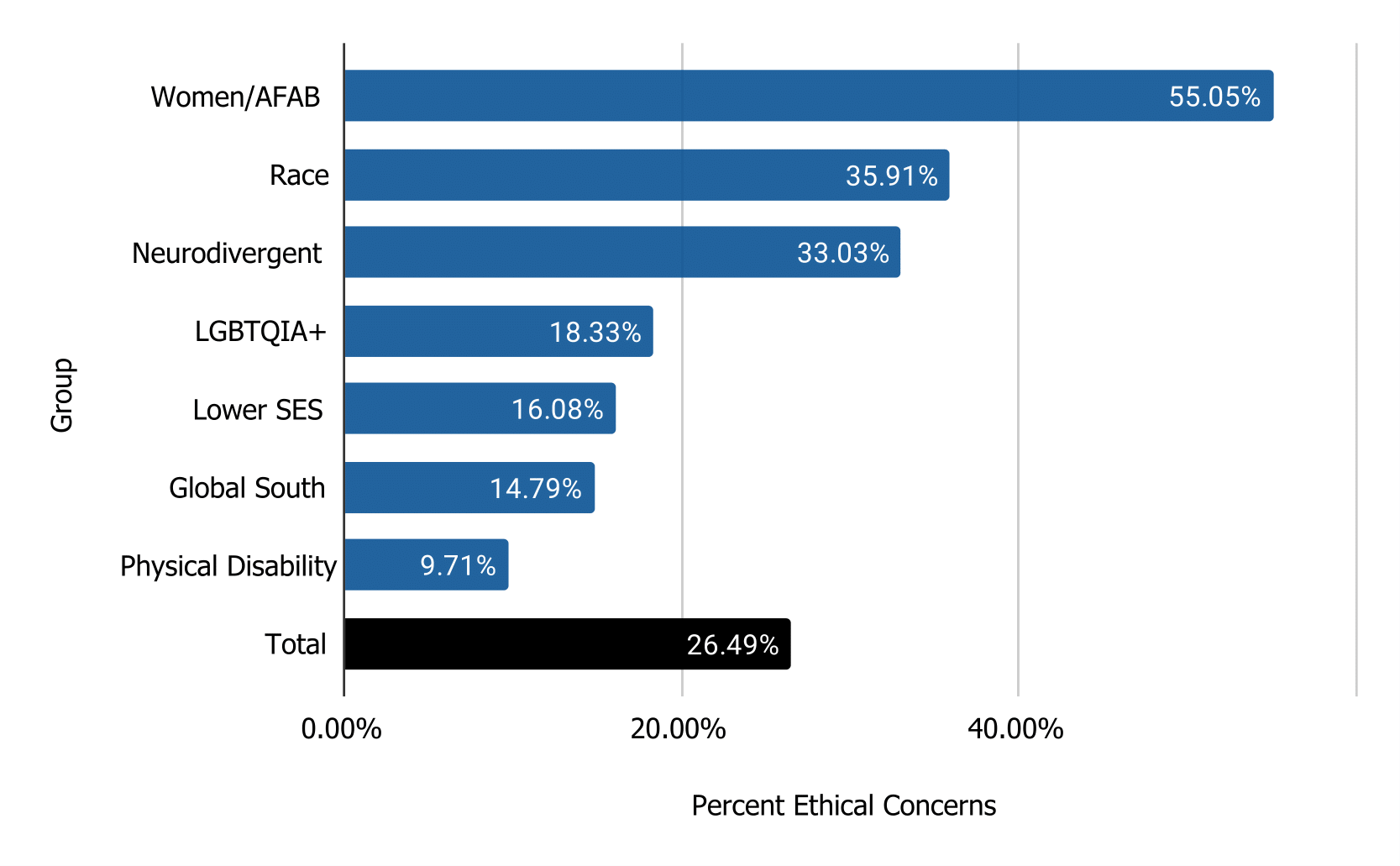}}}%
    
    \label{fig:groups}
\end{figure}

The three communities with higher-than-average reporting of ethical concerns are women\slash AFAB, race, and neurodivergent. The marginalized community with the greatest percentage of ethical concerns within their platform mentions is women\slash AFAB, with 55.05\%, as seen in Figure \ref{fig:groups}. The majority of these complaints revolve around censorship on Reddit, as discussed in Section \ref{sec:censorship}. The next highest-reporting ethical concerns community was the race community, whose most frequent grievance was by far discrimination. They are followed by the neurodivergent community, for whom addictive design was seen most frequently as an ethical concern. 

The four communities who report ethical concerns at a less-than-average rate are the queer, lower SES, Global South, and physical disability communities. The queer community's top ethical concern was also discrimination, with 14 reports of transphobia and homophobia, including a report of color under-representation in trans spaces. Next, the lower SES community reports cyberbullying and censorship as the most frequent concerns, with home-insecure users being harassed online and struggling in their dependence on platforms and random people online to provide them with resources needed for survival. The Global South's top ethical concerns were the other and scam categories, detailed further in Sections \ref{sec:scam} and \ref{sec:other}. The community with the least amount of ethical concerns proportionally, with 9.71\%, is the physical disability community, which was often positive about the connection that online spaces brought them, yet still had many claims of cyberbullying and medical misinformation.

%\subsubsection{Time}
 %In Figure \ref{fig:time}, it's clear where the bulk of the ethical concerns in our dataset originate from. The increase from 2012 to 2015 is due to complaints leveled against UserX and his censorship of feminist spaces on Reddit, detailed further in Section ~\ref{sec:censorship}. Otherwise, we can also see a slight increase starting after 2016, which could be attributed to recent popular ethics scandals on platforms. 

%\begin{figure}[h]
 %   \caption{Frequency of Ethical Concerns Over Time}

  %  \centerline{{\includegraphics[width=.5\textwidth]{figures/time.pdf}}}%
    
   % \label{fig:time}
%\end{figure}

\subsubsection{Platforms}
\label{sec:platforms}

\begin{figure}[h!]
    \caption{Ethical Concern Frequency by Platform}

    \centerline{{\includegraphics[width=.5\textwidth]{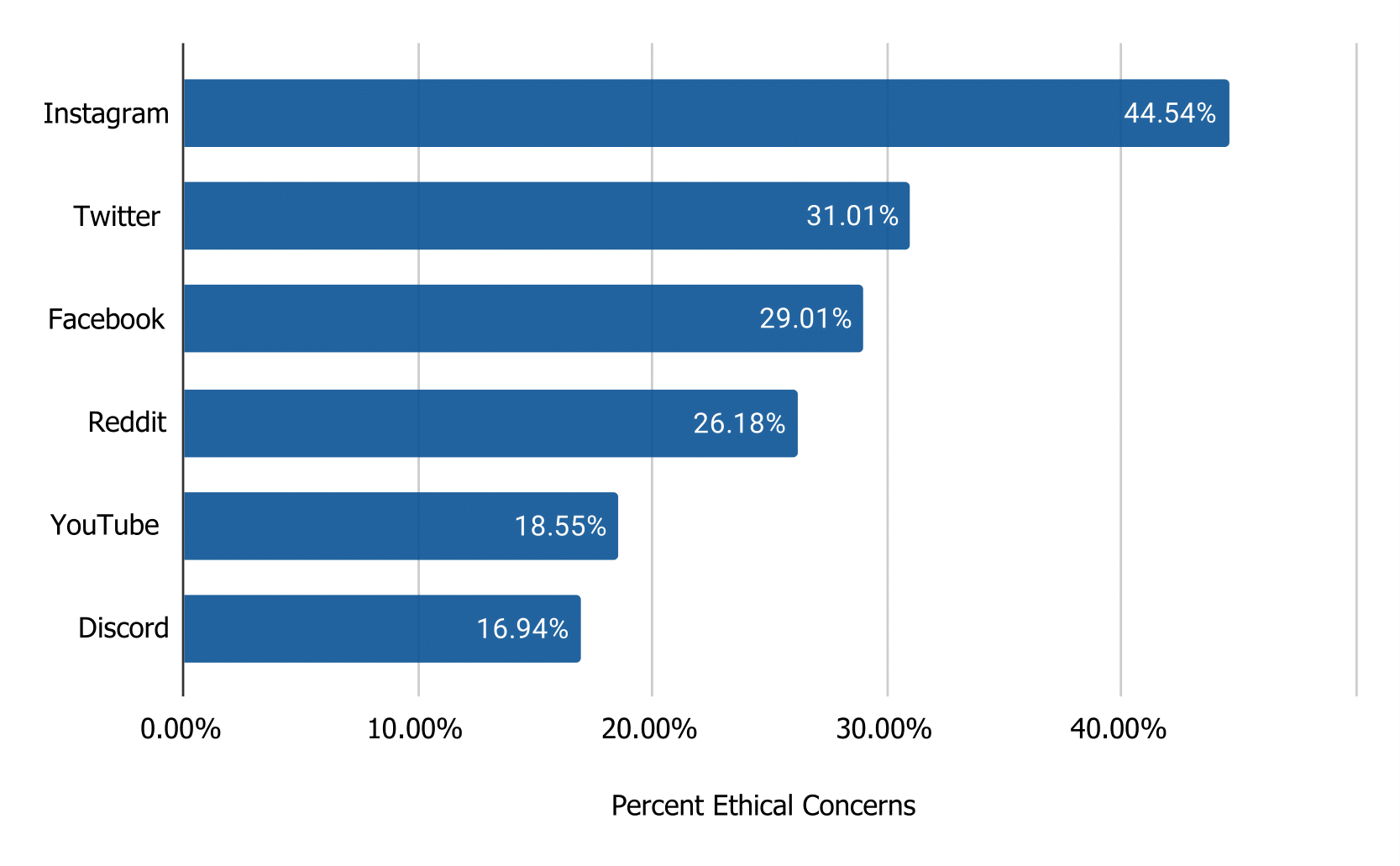}}}%
    
    \label{fig:platforms}
\end{figure}
Instagram, Facebook, and Discord all reported ethical concerns at a higher than average rate, while Reddit, Twitter, and YouTube are related to ethical concerns at a less than average rate, as seen in Figure \ref{fig:platforms}. To perform this analysis, we first removed platforms with less than a hundred total posts in our manually annotated dataset. Additionally, we removed concerns identified less than ten times from consideration. Finally, we normalize the frequency of ethical concerns by overall ethical concern type occurrence and the number of total ethical concerns on the platform in question. 

Instagram's highly occurring concerns are discrimination (28.30\%), inappropriate content (15.09\%), and scam (15.09\%), described further in Sections \ref{sec:discrimination}, \ref{sec:ic}, \ref{sec:scam}, respectively. For Facebook, the two most highly occurring ethical concerns are misinformation (10.53\%) and scam (15.79\%). The misinformation concerns relate to self-diagnoses for mental health concerns and stereotyping of lower SES and Asian people. Lastly, Discord users report social isolation (38.10\%) and cyberbullying (28.57\%) at the highest rate, signaling Discord's role in fostering relationships between users. 

Reddit's two highest concerns are censorship (42.32\%) and discrimination (22.87\%). Reddit's greatest ethical concerns have affected the frequency of ethical concerns we report in this paper as we performed it on Reddit and therefore elicited the most feedback on Reddit. Next, Twitter's highest concerns are cyberbullying (17.50\%) and inappropriate content (12.50\%). Twitter's cyberbullying concern describes constant toxic negativity on the platform and racism and lesbophobia. Again, their inappropriate content concern recounts negative content causing harm to users' mental health. Finally, YouTube's most frequent concerns were addictive design (24.39\%) and misinformation (12.2\%), described further in Sections \ref{sec:addict}, and \ref{sec:misinformation}, respectively.

\subsection{Type of Ethical Concerns Feedback}
\label{sec:content}
\begin{figure}
    \caption{Ethical Concern Frequency by Type}

    \centerline{{\includegraphics[width=.5\textwidth]{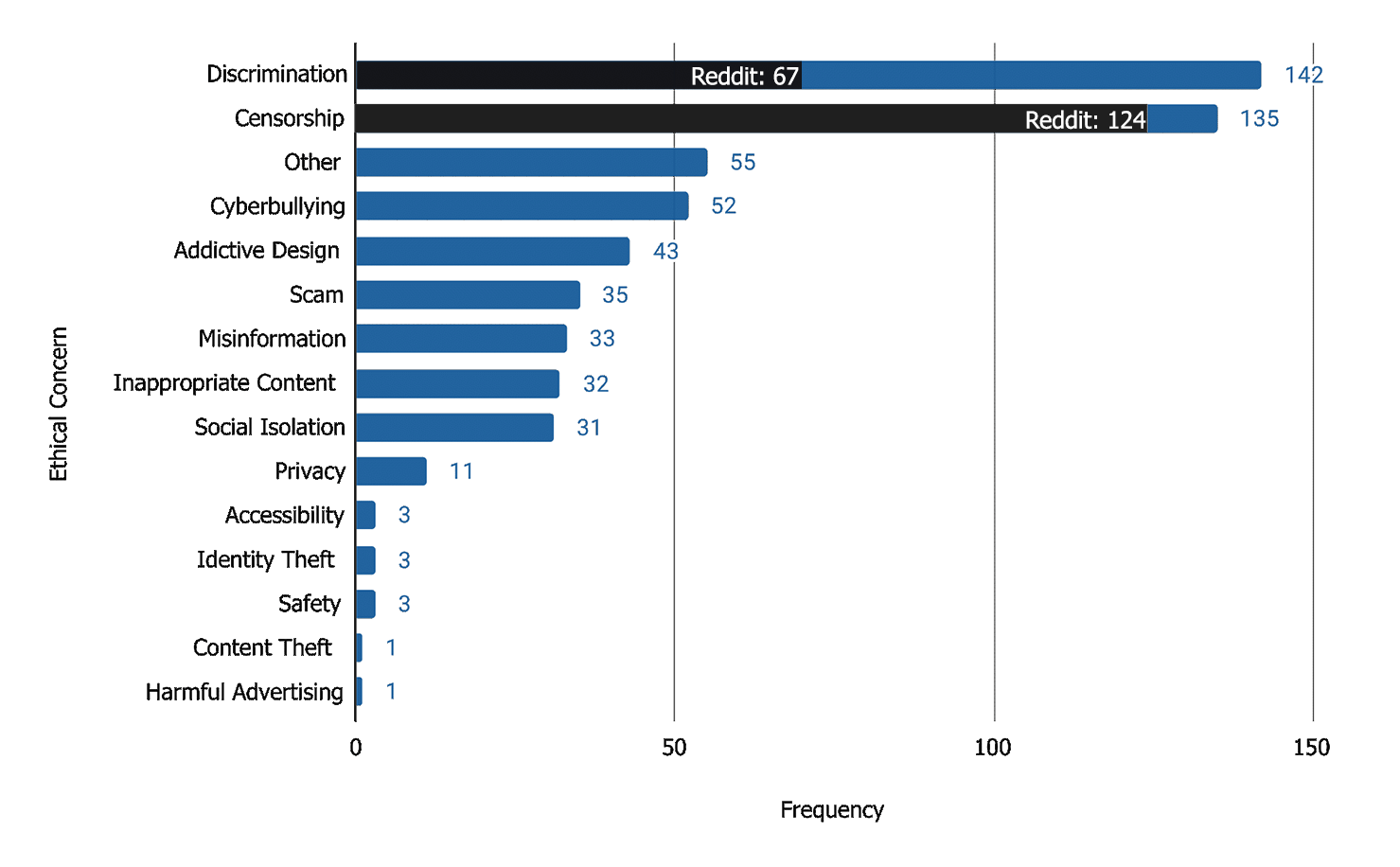}}}%
    
    \label{fig:types}
\end{figure}

In this section, we discuss each ethical concern type in detail. However, as referenced in the previous section, our frequency of types in Figure \ref{fig:types} are biased by the imbalance of platform posts, with Reddit (293) having over three times the number of posts in our ethical concerns as the next largest platform, YouTube (82), and nearly 100 times the amount of posts as our smallest platform, Snapchat (3). Because of this, we do not consider Figure \ref{fig:types} as representative of the actual distribution of concerns voiced by marginalized communities on social platforms. Instead, we examine each type of concern individually and report which marginalized communities they most frequently originate from, to treat these communities polylithically.

\subsubsection{Scam}
\label{sec:scam}
For the scam category, 78\%, or 28 of the 36 claims, were from the Global South or race category. Most of these claims detail influencers on YouTube and Instagram disingenuously endorsing and advertising cosmetic products to earn financial compensation. All of these claims originated from the ``r/indianskincareaddict," ``r/beautytalkph," and ``r/malaysianpf" subreddits. Due to the pressures of colorism faced by South Asian women/AFAB, it is no wonder that they face additional targeting from advertisers through influencers to buy skincare products~\cite{mishra2015india},\cite{norwood2013ubiquitousness}. %South Asian women/AFAB, among other women/AFAB of color, often face additional unrealistic beauty standards like skin lightening. 

\subsubsection{Misinformation}
\label{sec:misinformation}
Out of the 31 ethical concerns about misinformation, 20, or 64.5\%, relate to medical misinformation. Most of this medical misinformation originates from physical disability and neurodivergent-related subreddits. Half of these medical misinformation claims relate to material on YouTube. 

Based on the descriptions of users, there is a negative feedback loop where users need medical information and fall into an addictive pursuit of information relating to their condition in pursuit of a diagnosis. This medical information is often provided by non-medical professionals, diluting the reliability and quality of diagnosis and treatment. 

In addition to medical misinformation, users also describe misinformation for newly homeless people. For example, according to a post, telling newly homeless users to ``[c]all the cops" or apply for Section 8 can be dangerous as police often act violently towards homeless users, and the waiting list for Section 8 can be over 15 years long. 

\subsubsection{Inappropriate Content}
\label{sec:ic}
Within the inappropriate content concern, 50\% of the claims originate from the neurodivergent communities, with the majority of these posts describing anxiety or depression induced from online content. Oftentimes, unrealistic standards, popularized and elevated by ``the algorithm" to increase clicks, engender this negative mental reaction. The next largest group of users reporting inappropriate content is women/AFAB. Some women/AFAB feel inundated by wedding and fashion content, either feeling unable to live up to the standard, or trapped by their family's expectations for them. Otherwise, women/AFAB report misogyny online making them feel subhuman.

\subsubsection{Cyberbullying}

We address cyberbullying as a more user-driven form of inappropriate content; however, both are forms of content which harms users. 

First, it is critical to address the fetishization faced by women/AFAB of color and trans people online. Perpetrators post nonconsensual pornography to fetishize bodies and their connected identities. In our dataset, exclusively from the ``r/twoxindia" subreddit, a subreddit for Indian women/AFAB, there are six reports of nonconsensual pornography, the majority of this pornography hosted on Reddit. 

A difficulty in forming relationships online for trans users is their exposure to ``chasers," people who fetishize trans bodies. According to some users, it can be difficult to determine whether connections with other users online are genuine or potential harassment.

\subsubsection{Addictive design}
\label{sec:addict}
For the addictive design category, over 80\% of the claims originate from the neurodivergent community, suggesting that those with mental health struggles may be more vulnerable to online addiction. Of these, half of the complaints derive from the ADHD subreddit. Additionally, nearly 50\% of these claims concern YouTube. As many mental health disorders often co-occur with addictive behaviors, we can conclude it is likely that addictive design targets the mentally ill. Many of these claims describe platform use as a means to escape from and avoid their real-world problems.

\subsubsection{Other Ethical Concerns}
\label{sec:other}
A common concern of users from the Global South subreddits is the lack of required information that is available online. For example, when looking for information on politics and finance, users in the Global South often can only find information relating to the US rather than their own home countries. This lack of information is part of a broader trend of misrepresentation online expressed by all groups, whether it simply concerns false representation or the over or under-representation of certain groups. 

According to users, there are both instances of under and overrepresentation online, with women\slash AFAB reporting difficulty finding other women\slash AFAB users on Reddit and some users from our race group describing overrepresentation of popular members of their racial group on Twitter. 

 Another trend in misrepresentation is the assumption of the privileged group as default. On LinkedIn, some users describe receiving messages which start with ``Hello Mr. \_\_\_\_," despite not being a man. On Pinterest, multiple users complain that fashion-related images only feature white women, despite explicit searches for non-white women.
 
 \subsubsection{Censorship}
\label{sec:censorship}
For the censorship category, the most significant concern is voiced in response to the moderation of ``r\slash Feminism" by a male moderator, whom we will refer to as \textit{userX}. Most users within this concern claim \textit{userX} removed their posts and banned them after posting about common feminist issues. 

Of the 135 censorship complaints, 106, 79\%, are about \textit{userX}. These claims started in 2012 and dropped off after 2015. However, \textit{userX}, a man, remains the head moderator of the ``r\slash Feminism" subreddit. According to the \\ ``r\slash WhereAreTheFeminists" subreddit, another feminist was banned from ``r\slash Feminism" as recently as June 2022. 

These posts raise the question of whether censorship is inevitable when a centralized group controls a marginalized community's discourse. A similar claim was made by ``r/askacaribbean" regarding a Caribbean Facebook group which has two representatives who are not from the Caribbean region. The poster claims, therefore, that ``they don't understand the local culture and politics."

 \subsubsection{Social Isolation}
Nearly 47\% of social isolation concerns emanate from the neurodivergent group, with the following largest proportion, 27\%, arising from the queer community. The neurodivergent group reports exclusion and negativity online, two principles integrated into the structure of today's internet, with non-physical connections making it easier to neglect relationships and clickbait overpowering quality. 

According to the queer communities' posts, it can be unsafe to connect with other queer users in real life due to the abuse and harassment many queer people face. As a result, many can connect online to users facing similar struggles. However, these users still report harmful inundation and inauthenticity within online relationships.

\subsubsection{Privacy}
The privacy concerns originate mainly from the neurodivergent and queer communities. The mental health-related communities report online stalking from ex-partners and abusers and feel uncomfortable forming online relationships with people they know in real life. In the queer community, posts discuss whether and to what degree they should reveal their queer status. Giving more significant control over users' privacy from their own online and offline connections seems to be a safety issue, especially for women/AFAB, neurodivergent, and queer users facing potential stalking and abuse.

\subsubsection{Discrimination}
\label{sec:discrimination}
Within the discrimination concern, 57\% of claims originate from the race group. Of these claims, 33\% focus on inter and intra-group conflicts, specifically between races and genders within a race. Racism from white users is more pervasive, with 53/79, or 67\%, of reports of racism towards Asian or black users. The reports of discrimination from ``r/asianmasculinity," ``r/aznidentity," and ``r/southasianmasculinty" often mention online communities and platforms not disavowing racism and thus invalidating racism against Asian users. As discussed in Section \ref{sec:ic}, content online for women/AFAB of color, like those in ``r/blackgirldiaries," targets, dehumanizes, and reminds them of past trauma. This systematic exposure to degrading content leads to an inequitable experience online. Another trend among queer users, people of color, and those from the Global South is the invalidation of their identities.

\subsection{Automated Extraction Potential}
\label{sec:automatic}
Our binary classifier identifies ethical concerns within platform-mentioning posts, while our multi-class classifier discerns the type of the ethical concern. The intention of the automatic extraction of ethical concerns was to determine whether it is possible for software practitioners to aggregate and process this data on a large scale, to allow them to utilize these complaints within their software development process. Our top-performing binary and multi-class classifiers performs with f1-scores of .805 and .713, respectively.

\subsubsection{Classifiers}
 The Naive Bayes and SVM classifiers were chosen based on the recommendations from \cite{scikit} for predicting categories with labeled text data under 100k rows~.  The more specific instantiations of these models were chosen based on empirical validation. 

\begin{table}[h]
\caption{Binary Classification Results for Ethical Concern Presence}
\centering
\begin{tabular}{|p{0.2\textwidth} |c|c|c|}

\hline
 \textit{\textbf{Model}} & \textit{\textbf{Precision}} & \textit{\textbf{Recall}} & \textit{\textbf{F1-score}} \\ \hline
\textit{Logistic Regression} & 0.773 & 0.611 & 0.681 \\ \hline
\textit{Support Vector Machine} & \textbf{0.829} & 0.568 & 0.673  \\ \hline
\textit{LinearSVC} & 0.715 & 0.645 & 0.674 \\ \hline
\textit{NuSVC} & 0.812 & 0.800 & \textbf{0.805} \\ \hline
\textit{Gaussian Naive Bayes} & 0.731 & \textbf{0.889} & 0.802 \\ \hline

\end{tabular}

\label{table:results_bin}
\end{table}

\subsubsection{Binary Classification for Ethical Concerns}
These classifiers identify whether a platform-mentioning post contains an ethical concern or not.  Table~\ref{table:results_bin} shows the main results for each classifier. For this task, the Support Vector Machine had the highest precision score at 0.829. The Gaussian Naive Bayes classifier had the highest recall score at 0.889. Finally, the NuSVC classifier had the highest f1-score at 0.805.

\subsubsection{Multi-Class Classification for Ethical Concerns}
These classifiers identify the post's ethical concern type. For this task, we removed ethical concerns with less than 40 posts from the dataset used for prediction, only leaving the top 5 most frequent ethical concerns in our dataset. These concerns are discrimination, censorship, other, cyberbullying, and addictive design. Table~\ref{table:results_bin} shows the main results for each classifier. In all measures, the Gaussian Naive Bayes classifier was the most highly performing for this task, with precision=0.743, recall=0.725, and f1-score=0.713. When considering each individual ethical concern's ability to be categorized, censorship, across all models has the best performance. 

\begin{table}[h]
\caption{Multi-Class Classification Results for Ethical Concern Presence}
\centering
\begin{tabular}{|c|c|c|c|}

\hline
 \textit{\textbf{Model}} & \textit{\textbf{Precision}} & \textit{\textbf{Recall}} & \textit{\textbf{F1-score}} \\ \hline
\textit{Logistic Regression} & 0.739 & 0.647 & 0.666 \\ \hline
\textit{Support Vector Machine} & 0.694 & 0.591 & 0.597  \\ \hline
\textit{LinearSVC} & 0.695 & 0.690 & 0.684 \\ \hline
\textit{NuSVC} & 0.621 & 0.598 & 0.579 \\ \hline
\textit{Gaussian Naive Bayes} & \textbf{0.743} & \textbf{0.725} & \textbf{0.713} \\ \hline
\end{tabular}

\mycaption{\textit{Precision, Recall, and F1-scores are all macro-averaged}}
\label{table:results_multi}
\end{table}

\begin{table}[h]
\caption{Multi-Class Classification Results for Individual Labels}
\centering
\begin{tabular}{|c|c|c|c|c|}

\hline
 \textit{\textbf{}} & \textit{\textbf{LR}} & \textit{\textbf{SVM}} & \textit{\textbf{LSVC}} & \textbf{GNB}\\ \hline
\textit{Addictive Design} & 0.402 & 0.188 & 0.485 & 0.738 \\ \hline
\textit{Censorship} & \textbf{0.868} & \textbf{0.857} & \textbf{0.868} & \textbf{0.809} \\ \hline
\textit{Cyberbullying} & 0.210 & 0.130 & 0.086 & 0.766 \\ \hline
\textit{Discrimination} & 0.672 & 0.636 & 0.649 & 0.633 \\ \hline
\textit{Other Ethical Concerns} & 0.151 & 0.095 & 0.067 & 0.590 \\ \hline
\end{tabular}

\label{table:results_multi_labels}
\end{table}
\section{Discussion}
Our results demonstrate that (\textbf{\textit{RQ1}}) Reddit is a valuable source for feedback on ethical concerns from marginalized communities, (\textbf{\textit{RQ2}}) marginalized communities’ ethical concerns center around manipulation and oppression, and (\textbf{\textit{RQ3}}) there is potential for this data to be collected and classified on a large scale, to collate ethical concerns.

 To address Reddit’s potential as a source of marginalized communities’ ethical concerns on software (\textbf{\textit{RQ1}}), we report that marginalized communities’ subreddits are a potent source of feedback on ethical concerns. Despite only a 5\% platform-mention rate in marginalized users’ posts, the substantial amount of Reddit data still leaves us plenty of content for manual and automated analysis. Additionally, the level of ethical concerns is high, with 26.49\% of platform-mentioning posts referencing an ethical concern, comparable to Obie et al's rate of 26.5\% human values violations found in user feedback\cite{obie2021first}. However, although Reddit generally contains a high rate of ethical concerns about software, Reddit’s efficacy shifts for different communities. For example, women/AFAB (55.05\%) had a far higher percentage of ethical concerns than physically disabled users (9.71\%). Additionally, we found that our posts with ethical concerns were on average 1906.8 characters long. This high average exemplifies Reddit's ability to facilitate contextualizing detail in its posts through its high character limit. Within posts, there is often detail on why and how these ethical concerns arise as well as the effect of these ethically-concerning situations on users. 

Next, to address marginalized communities’ ethical concerns (\textbf{\textit{RQ2}}), we discuss the manipulation and oppression faced by users online. By structuring platform development around profit, platforms have placed undue burdens on marginalized communities. Through manual analysis of posts, we discovered that there seems to be a tentative relationship between inappropriate content, cyberbullying, scams, misinformation, and addictive design. In this relationship, platforms elevate inappropriate and misleading information due to its volatile nature, making it addictive to consume. This phenomenon allows platforms and content creators to profit via advertising and manipulative practices \cite{gray2018dark}. Our data shows that online manipulation and harassment may target the most vulnerable and desperate for resources. For example, medical misinformation targets users who have stigmatized medical problems like those relating to mental health or sex. To assuage some online manipulation, platforms could add stricter regulations for advertisers and harsher rules regarding scams by independent creators. Moreover, they could try to identify and aid potentially targeted vulnerable populations. 

Another example of vulnerable groups being targeted is women/AFAB, especially women/AFAB of color. It can be challenging to remove nonconsensual pornography from platforms, with one user describing a situation where a 16-year-old’s naked pictures were posted online. However, she could not file a police report out of fear that her parents would find out. According to another user, women/AFAB’s fully-clothed photos uploaded from Eid were used to “auction...[women/AFAB] on live stream on youtube.” Despite their lack of nudity, they were still being sexualized and rated~\cite{khan_2022}. This sexualization points to the issue not being with women/AFAB and what they decide to do with their bodies but how those bodies are perceived and treated. As such, developers should change how they deal with nonconsensual pornography to remove the burden from the victims and put it back on the perpetrators. 

Finally, a frequent narrative in the zeitgeist seems to be that \textit{everything} can be found online. However, this is certainly not the case for all users, who lack basic means and may depend more heavily on the Internet for resources and information. Despite this greater need, marginalized communities often find inaccurate information or information geared toward centralized groups rather than marginalized ones. Like Wikipedia, other software platforms could devote more resources to equitable and objective informational resources. 

In addition, as relationships and identities shift online, software engineers create their structures, evaluatory metrics, and interaction capabilities. According to users, these tools often fall short and provide connections once impossible. Platforms should reassess privacy threats by considering how marginalized communities may need to shift their identity expression contextually for safety purposes. Furthermore, they could create design features that mimic real-life social interaction more closely to help users build healthier relationships. 

In Simone de Beauvoir’s \textit{The Second Sex}, she argues that the man is regarded as the default, while the woman is regarded as the “Other;”\cite{beauvoir_2015} unfortunately, this phenomenon extends beyond gender into other marginalized identities. Through the development of online identities and relationships, the “Other[ing]” and further misrepresentation of marginalized users seems to follow. Under-representation of an identity's full diversity and over-representation of a small sample of an identity lead to the misrepresentation of these identities. As a result, pervasive stereotyping simplifies their identities to often distorted and inaccurate perceptions. Creating a democratic process for electing moderators could help solve the control of marginalized communities’ online spaces by their more privileged counterparts to allow these groups to represent themselves. By making sure to elevate greater amounts of marginalized group members, platforms could help more accurately represent their complexity. Additionally, stopping pre-selection for and assumption of privileged identities in forms and other areas may help decouple marginalized communities’ representation as “Other” from design.

To address the discrimination in content moderation, we discuss the inter and intra-group conflict found in our posts. The cause of this conflict may be what Du Bois defines as a “double consciousness,” where marginalized communities see themselves through the view of the privileged group in addition to their own. Marginalized communities adopt the harmful centralized perspectives of, in this case, their and others’ races and treat themselves and others with this viewpoint\cite{du2015souls}. Intra-group conflict is also frequent in trans spaces online, where trans users also deal with a “double consciousness,” which derives from the survivalist need to fit into cisheteronormative society and the right to be one’s actual self. This friction also leads to conflict over how to define and qualify the trans experience. Based on the posts, this conflict may also partly arise from intergenerational conflict, where queer users feel this “double consciousness” at varying levels based on when and where they were born. Currently, content moderation algorithms over-police non-white~\cite{davidson2019racial} and queer communities~\cite{dias2021fighting} because of these communities’ use of their vernaculars. However, examining instances of differing inter and intra-group language types reported by users in our posts and developing content moderation techniques centered around marginalized users’ perspectives could make these algorithms more just. 

Finally, concerning the tasks’ automation potential (\textbf{\textit{RQ3}}), we attempt to explain our relatively high F1-scores, considering the different manifestations and words by which the same ethical concerns are described. First, our experiment is lightly comparable to Iqbal et al.’s work,\cite{iqbal2021mining}, which identifies user feedback, like bug reports and feature requests, from platforms’ subreddits on Reddit. Like us, they report on performance in the [.7, .9] range for binary classifiers. Both our high scores suggest that developers could collect this data on a larger scale to identify gaps in current feedback elicitation practices. Also, our Gaussian Naive Bayes classifier has a recall of .88. Recall is a more critical measure for this binary task, as we would like to collect all possible ethical concerns rather than leave some undiscovered. Additionally, our multi-class classifiers categorize ethical concern type with an F1-score of .743. Although, we can surmise that the censorship category performed most highly likely due to its high occurrence (84.5\%) among a single subreddit, ``r/WhereAreTheFeminists". With a more variable dataset, the performance of the multi-class classifiers would drop. Although, a fully classified dataset could give developers even more insight into correlations between ethical concerns’ types, occurrence on platforms, and different marginalized communities. 
Although we do not see our current model as highly generalizable due to our small dataset and limited selection of subreddits, a more generalizable model could give fine-grained insight into the 586 different marginalized communities we targeted in our study, among other potential marginalized communities.

\section{Threats to Validity}
A base shortcoming of this paper is the demographics of the authors. While each community is represented in some manner by one of the authors, our only POC author is white-passing, which may weaken our discussion of colorism and racism online users face. Furthermore, we do not represent each of the 586 different communities and did not include these users in the analysis of their concerns. 

Another weakness of this paper is our ability to handle feedback from the Global South. We only annotated posts in English, which creates a language barrier between our work and non-native English speakers. Again, as we desired to capture marginalized feedback, this is a crucial weakness. Future work could analyze similar posts in other languages. 

Additionally, Reddit tends to have the exact demographic we are looking not to represent. According to Statista, most Reddit users are white, American men~\cite{dixon_2022},\cite{dixon_2016},\cite{reddit_2022}. However, we attempted to mitigate this issue by explicitly focusing on subreddits that originated from marginalized communities rather than from platforms or regarding technology. Unfortunately, this demographic inequality and our use of a popularity metric for subreddit selection may have biased group selection towards men. For example, the analyzed subreddits ``r\slash malementalhealth," ``r\slash asianmasculinity," ``r\slash southasianmasculinity," and ``r\slash prostatitis" are geared toward men.

Future work could also handle intersectionality more adeptly than this paper. For example, we have multiple intersectional subreddits. Although, for simplification's sake, we had to categorize subreddits into just one identity based on the time of discovery. So, for example, ``r\slash twoxindia" was included in the women\slash AFAB category rather than the Global South category. 
Future work could also endeavor to exclude Reddit as a social platform from the filtering process as we posit that our results are skewed by our data source's influence. More generally, our results are highly dependent on the chosen platforms and future work should seek to eliminate this bias; however, these platforms are ubiquitous amongst users.  

It is critical to note that in this paper, not every ethical concern detailed in our dataset was analyzed and discussed. Even the ethical concerns that made it into this paper could likely generate their own research directions. Beyond the labeled data that was left out, and although we scraped 586 groups' subreddits, less than a tenth of those subreddits made it to the data annotation process. Although we posit that the concerns described in this paper are general, critical phenomena faced by billions of users, these concerns do not affect every user equally. 

Considering our machine learning models, we cannot conclude that this model is generalizable to a larger dataset when the data we trained the model on is specific to a smaller group of subreddits. Although as is the trend within software engineering, we cannot create a model that fits all users, all people. Instead of insisting on creating top-down approaches to software design, we ought to shift, more generally, to bottom-up approaches, which are founded on the concerns of different groups and end in alternate products and experiences.

\section{Conclusion}
There is a long history of corporations exploiting consumers and privileged groups benefiting from the abuse of marginalized communities. Unfortunately, this trend is now mirrored in the transition to the online world. Though, with it, has come a democratization of access and expression. Importantly, making user feedback actionable, and doing so with a prioritization of those along the margins, gives those oppressed potential to change the software which affects them deeply. As the development process becomes more agile, software can change quickly, and the views of those often overlooked can drive those updates. Our work contributes to this pursuit by collating and analysing marginalized communities' ethical concerns in  mentions of platforms in Reddit posts.

%\section*{Acknowledgment}
%sorry mom

\bibliographystyle{IEEEtran}
\bibliography{citations}

\begin{thebibliography}{10}
\providecommand{\url}[1]{#1}
\csname url@samestyle\endcsname
\providecommand{\newblock}{\relax}
\providecommand{\bibinfo}[2]{#2}
\providecommand{\BIBentrySTDinterwordspacing}{\spaceskip=0pt\relax}
\providecommand{\BIBentryALTinterwordstretchfactor}{4}
\providecommand{\BIBentryALTinterwordspacing}{\spaceskip=\fontdimen2\font plus
\BIBentryALTinterwordstretchfactor\fontdimen3\font minus
  \fontdimen4\font\relax}
\providecommand{\BIBforeignlanguage}[2]{{%
\expandafter\ifx\csname l@#1\endcsname\relax
\typeout{** WARNING: IEEEtran.bst: No hyphenation pattern has been}%
\typeout{** loaded for the language `#1'. Using the pattern for}%
\typeout{** the default language instead.}%
\else
\language=\csname l@#1\endcsname
\fi
#2}}
\providecommand{\BIBdecl}{\relax}
\BIBdecl

\bibitem{biddle_2022}
\BIBentryALTinterwordspacing
S.~Biddle, ``Facebook report concludes company censorship violated palestinian
  human rights,'' Sep 2022. [Online]. Available:
  \url{https://theintercept.com/2022/09/21/facebook-censorship-palestine-israel-algorithm/}
\BIBentrySTDinterwordspacing

\bibitem{rohingya_2022}
\BIBentryALTinterwordspacing
``Myanmar: The social atrocity: Meta and the right to remedy for the
  rohingya,'' Sep 2022. [Online]. Available:
  \url{https://www.amnesty.org/en/documents/ASA16/5933/2022/en/}
\BIBentrySTDinterwordspacing

\bibitem{chayka_2022}
\BIBentryALTinterwordspacing
K.~Chayka, ``The online spaces that enable mass shooters,'' May 2022. [Online].
  Available:
  \url{https://www.newyorker.com/culture/infinite-scroll/the-online-spaces-that-enable-mass-shooters}
\BIBentrySTDinterwordspacing

\bibitem{woolf_2014}
\BIBentryALTinterwordspacing
N.~Woolf, ``'puahate' and 'foreveralone': Inside elliot rodger's online life,''
  May 2014. [Online]. Available:
  \url{https://www.theguardian.com/world/2014/may/30/elliot-rodger-puahate-forever-alone-reddit-forums}
\BIBentrySTDinterwordspacing

\bibitem{digital_hate2022}
\BIBentryALTinterwordspacing
``Digital hate - hrc-prod-requests.s3-us-west-2.amazonaws.com.'' [Online].
  Available:
  \url{https://hrc-prod-requests.s3-us-west-2.amazonaws.com/CCDH-HRC-Digital-Hate-Report-2022-single-pages.pdf}
\BIBentrySTDinterwordspacing

\bibitem{tiffany_2020}
\BIBentryALTinterwordspacing
K.~Tiffany, ``The women making conspiracy theories beautiful,'' Aug 2020.
  [Online]. Available:
  \url{https://www.theatlantic.com/technology/archive/2020/08/how-instagram-aesthetics-repackage-qanon/615364/}
\BIBentrySTDinterwordspacing

\bibitem{costanza2020design}
S.~Costanza-Chock, \emph{Design justice: Community-led practices to build the
  worlds we need}.\hskip 1em plus 0.5em minus 0.4em\relax The MIT Press, 2020.

\bibitem{tizard2020voice}
J.~Tizard, T.~Rietz, and K.~Blincoe, ``Voice of the users: A demographic study
  of software feedback behaviour,'' in \emph{2020 IEEE 28th International
  Requirements Engineering Conference (RE)}.\hskip 1em plus 0.5em minus
  0.4em\relax IEEE, 2020, pp. 55--65.

\bibitem{gender_user}
E.~Guzman and A.~Paredes~Rojas, ``Gender and user feedback: An exploratory
  study,'' in \emph{2019 IEEE 27th International Requirements Engineering
  Conference (RE)}, 2019, pp. 381--385.

\bibitem{same_same}
E.~Oehri and E.~Guzman, ``Same same but different: Finding similar user
  feedback across multiple platforms and languages,'' in \emph{2020 IEEE 28th
  International Requirements Engineering Conference (RE)}, 2020, pp. 44--54.

\bibitem{user_cross_cultural}
E.~Guzman, L.~Oliveira, Y.~Steiner, L.~C. Wagner, and M.~Glinz, ``User feedback
  in the app store: A cross-cultural study,'' in \emph{2018 IEEE/ACM 40th
  International Conference on Software Engineering: Software Engineering in
  Society (ICSE-SEIS)}, 2018, pp. 13--22.

\bibitem{context_martens}
D.~Martens and W.~Maalej, ``Extracting and analyzing context information in
  user-support conversations on twitter,'' in \emph{2019 IEEE 27th
  International Requirements Engineering Conference (RE)}, 2019, pp. 131--141.

\bibitem{workmanfront}
H.~Workman and C.~A. Coleman, ``“the front page of the internet”: Safe
  spaces and hyperpersonal communication among females in an online
  community,'' \emph{Southwestern Mass Communication Journal}, vol.~29, no.~2,
  2014.

\bibitem{o2018today}
T.~O'Neill, ``‘today i speak’: Exploring how victim-survivors use reddit,''
  \emph{International journal for crime, justice and social democracy}, vol.~7,
  no.~1, p.~44, 2018.

\bibitem{leavitt2015throwaway}
A.~Leavitt, ``" this is a throwaway account" temporary technical identities and
  perceptions of anonymity in a massive online community,'' in
  \emph{Proceedings of the 18th ACM conference on computer supported
  cooperative work \& social computing}, 2015, pp. 317--327.

\bibitem{koepfler2013stake}
J.~A. Koepfler, K.~Shilton, and K.~R. Fleischmann, ``A stake in the issue of
  homelessness: Identifying values of interest for design in online
  communities,'' in \emph{Proceedings of the 6th International Conference on
  Communities and Technologies}, 2013, pp. 36--45.

\bibitem{rankin2021resisting}
Y.~A. Rankin and K.~K. Henderson, ``Resisting racism in tech design: Centering
  the experiences of black youth,'' \emph{Proceedings of the ACM on
  Human-Computer Interaction}, vol.~5, no. CSCW1, pp. 1--32, 2021.

\bibitem{Pagano2013}
D.~Pagano and B.~Bruegge, ``{User Involvement in Software Evolution Practice :
  A Case Study},'' in \emph{Proc. of the International Conference on Software
  Engineering}, 2013, pp. 953--962.

\bibitem{groen2017crowd}
E.~C. Groen, N.~Seyff, R.~Ali, F.~Dalpiaz, J.~Doerr, E.~Guzman, M.~Hosseini,
  J.~Marco, M.~Oriol, A.~Perini \emph{et~al.}, ``The crowd in requirements
  engineering: The landscape and challenges,'' \emph{IEEE software}, vol.~34,
  no.~2, pp. 44--52, 2017.

\bibitem{Johann2015}
T.~Johann and W.~Maalej, ``{Democratic mass participation of users in
  Requirements Engineering?}'' in \emph{Proc. of the International Requirements
  Engineering Conference (RE)}, aug 2015, pp. 256--261.

\bibitem{Dennis2013}
D.~Pagano and W.~Maalej, ``{User feedback in the appstore: an empirical
  study},'' in \emph{Proc. of the International Requirements Engineering
  Conference}, 2013, pp. 125--134.

\bibitem{hoon2013analysis}
L.~Hoon, R.~Vasa, J.-G. Schneider, J.~Grundy, and Others, ``{An analysis of the
  mobile app review landscape: trends and implications},'' \emph{Swinburne
  University of Technology, Tech. Rep}, 2013.

\bibitem{Guzman2016}
E.~Guzman, R.~Alkadhi, and N.~Seyff, ``{A Needle in a Haystack: What Do Twitter
  Users Say about Software?}'' in \emph{Proc. of the International Requirements
  Engineering Conference}, 2016, pp. 96--105.

\bibitem{guzman2017little}
E.~Guzman, M.~Ibrahim, and M.~Glinz, ``A little bird told me: Mining tweets for
  requirements and software evolution,'' in \emph{Proc. of the International
  Requirements Engineering Conference (RE)}, 2017, pp. 11--20.

\bibitem{nayebi2018app}
M.~Nayebi, H.~Cho, and G.~Ruhe, ``App store mining is not enough for app
  improvement,'' \emph{Empirical Software Engineering}, vol.~23, no.~5, pp.
  2764--2794, 2018.

\bibitem{williams2017mining}
G.~Williams and A.~Mahmoud, ``Mining twitter feeds for software user
  requirements,'' in \emph{Proc. of the International Requirements Engineering
  Conference (RE)}, 2017, pp. 1--10.

\bibitem{iqbal2021mining}
T.~Iqbal, M.~Khan, K.~Taveter, and N.~Seyff, ``Mining reddit as a new source
  for software requirements,'' in \emph{2021 IEEE 29th International
  Requirements Engineering Conference (RE)}.\hskip 1em plus 0.5em minus
  0.4em\relax IEEE, 2021, pp. 128--138.

\bibitem{fischer2021does}
R.~A.-L. Fischer, R.~Walczuch, and E.~Guzman, ``Does culture matter? impact of
  individualism and uncertainty avoidance on app reviews,'' in \emph{2021
  IEEE/ACM 43rd International Conference on Software Engineering: Software
  Engineering in Society (ICSE-SEIS)}.\hskip 1em plus 0.5em minus 0.4em\relax
  IEEE, 2021, pp. 67--76.

\bibitem{tushev2020digital}
M.~Tushev, F.~Ebrahimi, and A.~Mahmoud, ``Digital discrimination in sharing
  economy a requirements engineering perspective,'' in \emph{2020 IEEE 28th
  International Requirements Engineering Conference (RE)}.\hskip 1em plus 0.5em
  minus 0.4em\relax IEEE, 2020, pp. 204--214.

\bibitem{besmer2020investigating}
A.~R. Besmer, J.~Watson, and M.~S. Banks, ``Investigating user perceptions of
  mobile app privacy: An analysis of user-submitted app reviews,''
  \emph{International Journal of Information Security and Privacy (IJISP)},
  vol.~14, no.~4, pp. 74--91, 2020.

\bibitem{reddit_privacy}
Z.~S. Li, M.~Sihag, N.~N. Arony, J.~B. Junior, T.~Phan, N.~Ernst, and
  D.~Damian, ``Narratives: the unforeseen influencer of privacy concerns,'' in
  \emph{2022 IEEE 30th International Requirements Engineering Conference (RE)},
  2022, pp. 127--139.

\bibitem{khalid2014mobile}
H.~Khalid, E.~Shihab, M.~Nagappan, and A.~E. Hassan, ``What do mobile app users
  complain about?'' \emph{IEEE software}, vol.~32, no.~3, pp. 70--77, 2014.

\bibitem{shams2020society}
R.~A. Shams, W.~Hussain, G.~Oliver, A.~Nurwidyantoro, H.~Perera, and
  J.~Whittle, ``Society-oriented applications development: Investigating
  users’ values from bangladeshi agriculture mobile applications,'' in
  \emph{2020 IEEE/ACM 42nd International Conference on Software Engineering:
  Software Engineering in Society (ICSE-SEIS)}.\hskip 1em plus 0.5em minus
  0.4em\relax IEEE, 2020, pp. 53--62.

\bibitem{obie2021first}
H.~O. Obie, W.~Hussain, X.~Xia, J.~Grundy, L.~Li, B.~Turhan, J.~Whittle, and
  M.~Shahin, ``A first look at human values-violation in app reviews,'' in
  \emph{2021 IEEE/ACM 43rd International Conference on Software Engineering:
  Software Engineering in Society (ICSE-SEIS)}.\hskip 1em plus 0.5em minus
  0.4em\relax IEEE, 2021, pp. 29--38.

\bibitem{schwartz2012overview}
S.~H. Schwartz, ``An overview of the schwartz theory of basic values,''
  \emph{Online readings in Psychology and Culture}, vol.~2, no.~1, pp.
  2307--0919, 2012.

\bibitem{ethical_concerns}
L.~Olson, N.~Tjikhoeri, and E.~Guzmán, ``The best ends for the best means:
  Ethical concerns in app reviews,'' 2024.

\bibitem{who}
\BIBentryALTinterwordspacing
``Disability and health.'' [Online]. Available:
  \url{https://www.who.int/news-room/fact-sheets/detail/disability-and-health}
\BIBentrySTDinterwordspacing

\bibitem{chawla2002smote}
N.~V. Chawla, K.~W. Bowyer, L.~O. Hall, and W.~P. Kegelmeyer, ``Smote:
  synthetic minority over-sampling technique,'' \emph{Journal of artificial
  intelligence research}, vol.~16, pp. 321--357, 2002.

\bibitem{mishra2015india}
N.~Mishra, ``India and colorism: The finer nuances,'' \emph{Wash. U. Global
  Stud. L. Rev.}, vol.~14, p. 725, 2015.

\bibitem{norwood2013ubiquitousness}
K.~J. Norwood and V.~S. Foreman, ``The ubiquitousness of colorism: Then and
  now,'' in \emph{Color Matters}.\hskip 1em plus 0.5em minus 0.4em\relax
  Routledge, 2013, pp. 9--28.

\bibitem{scikit}
\BIBentryALTinterwordspacing
``Choosing the right estimator.'' [Online]. Available:
  \url{https://scikit-learn.org/stable/tutorial/machine\_learning\_map/index.html}
\BIBentrySTDinterwordspacing

\bibitem{gray2018dark}
C.~M. Gray, Y.~Kou, B.~Battles, J.~Hoggatt, and A.~L. Toombs, ``The dark
  (patterns) side of ux design,'' in \emph{Proceedings of the 2018 CHI
  conference on human factors in computing systems}, 2018, pp. 1--14.

\bibitem{khan_2022}
\BIBentryALTinterwordspacing
A.~Khan, ``The invisible trauma of 'bhabhi' porn in the lives of indian
  women,'' Jun 2022. [Online]. Available:
  \url{https://www.vice.com/en/article/pkgdzg/bhabhi-porn-trauma-on-indian-women-savita-bhabhi-fetish-sex}
\BIBentrySTDinterwordspacing

\bibitem{beauvoir_2015}
S.~d. Beauvoir, \emph{The Second sex}.\hskip 1em plus 0.5em minus 0.4em\relax
  Vintage Classic, 2015.

\bibitem{du2015souls}
W.~E.~B. Du~Bois and M.~Marable, \emph{Souls of black folk}.\hskip 1em plus
  0.5em minus 0.4em\relax Routledge, 2015.

\bibitem{davidson2019racial}
T.~Davidson, D.~Bhattacharya, and I.~Weber, ``Racial bias in hate speech and
  abusive language detection datasets,'' \emph{arXiv preprint
  arXiv:1905.12516}, 2019.

\bibitem{dias2021fighting}
T.~Dias~Oliva, D.~M. Antonialli, and A.~Gomes, ``Fighting hate speech,
  silencing drag queens? artificial intelligence in content moderation and
  risks to lgbtq voices online,'' \emph{Sexuality \& Culture}, vol.~25, no.~2,
  pp. 700--732, 2021.

\bibitem{dixon_2022}
\BIBentryALTinterwordspacing
S.~Dixon, ``Global reddit user distribution by gender 2022,'' Mar 2022.
  [Online]. Available:
  \url{https://www.statista.com/statistics/1255182/distribution-of-users-on-reddit-worldwide-gender/}
\BIBentrySTDinterwordspacing

\bibitem{dixon_2016}
\BIBentryALTinterwordspacing
------, ``U.s. reddit user share by ethnicity 2016,'' Feb 2016. [Online].
  Available:
  \url{https://www.statista.com/statistics/517229/reddit-user-distribution-usa-ethnicity/}
\BIBentrySTDinterwordspacing

\bibitem{reddit_2022}
\BIBentryALTinterwordspacing
 [Online]. Available:
  \url{https://worldpopulationreview.com/country-rankings/reddit-users-by-country}
\BIBentrySTDinterwordspacing

\end{thebibliography}

\end{document}